\documentclass[aps,prd,reprint,twocolumn,letterpaper,superscriptaddress,tightenlines,nofootinbib,preprintnumbers]{revtex4-1}
\pdfoutput=1 
\usepackage{graphicx,dcolumn,bm,epsfig,cancel,hyperref,amsmath,amssymb,color}

\begin{document}
\title{Effective Theory and Simple Completions for Neutrino Interactions}
\author{Mark B. Wise}
\affiliation{California Institute of Technology, Pasadena, CA 91125}
\author{Yue Zhang}
\affiliation{California Institute of Technology, Pasadena, CA 91125}

\begin{abstract}
We consider all the dimension 6 operators as well as some simple extensions of the standard model that give new contributions to neutrino interactions with matter. Such interactions are usually parametrized by $\epsilon_{\alpha \beta}$, where $\alpha$ and $\beta$ are neutrino flavor indices taking the values $e$, $\mu$ and $\tau$. 
In the simple models we consider the $\epsilon_{\alpha \beta}$'s are much more constrained than in the operator-based model-independent approach. Typically the $\epsilon_{\alpha \beta}$'s are restricted to be smaller in magnitude than around $10^{-3}$. In some of the leptoquark models, a specific pattern for the leptoquark Yukawa couplings allows the diagonal element $\epsilon_{\tau\tau}$ to be as large as $\sim0.1$, or one of $\epsilon_{ee}$, $\epsilon_{\mu\mu}\sim0.01$. We discuss the interplay between neutrino physics  and leptoquark searches at the LHC.
\end{abstract}

\preprint{CALT-68-2884}

\maketitle

\section{Introduction}

There are many unanswered questions with regard to neutrinos.  Are their masses Majorana or Dirac?  What are the hierarchy and absolute scale of their masses?  Is there significant CP violation in the neutrino sector? Some of the experiments that bear on these questions are also sensitive to non-standard neutrino interactions with matter. Since the left handed neutrinos and the charged leptons are in $SU(2)_L$ doublets, new interactions for the neutrinos are very constrained by experiments, many of which do not involve the neutrinos at all. In this paper we consider non-standard interactions of neutrinos with matter~\cite{Ohlsson:2012kf}. 
One approach to beyond the standard model physics that contributes to neutrino interactions with matter is an operator analysis~\cite{Biggio:2009nt,Davidson:2003ha,Mohapatra:2005wg}. This approach has the advantage of being very general but can overlook correlations that occur in models. There is value in using  both the general operator analysis approach and the simple model approach to assess the likelihood that beyond the standard model interactions of neutrinos with matter can impact the future and present neutrino experimental program. 

In the standard model (SM) W boson exchange gives rise to the following effective Hamiltonian for neutrino interactions with ordinary matter
\begin{equation}
H_W= \sqrt{2}{G_F}[ {\bar \nu}_e \gamma^{\mu}P_L \nu_e ][{\bar e}\gamma^{\mu} e +\ldots] \ ,
\end{equation}
where $P_L=(1-\gamma_5)/2$ is the left-handed projection operator. Here we have ignored other terms like the axial couplings and the other leptons which do not play a role. For neutrino interactions with matter that has a number density of electrons $n_e$ this becomes,
\begin{equation}
H_W=  \sqrt{2}{G_F}[ {\bar \nu}_e \gamma^{0} P_L\nu_e ]n_e \ .
\end{equation}
Physics beyond the SM may contribute to the interactions of neutrinos with ordinary matter,  atoms composed of protons and neutrons and electrons.  Approximating $n_p = n_n = n_e = n$ these interactions have the form
\begin{equation}
\label{NSIdef}
H_{\rm BSM}=  \sqrt{2}{G_F}\sum_{\alpha,\beta=e,\mu,\tau}\epsilon_{\alpha \beta} [ {\bar \nu}_{\alpha} \gamma^{\mu}P_L \nu_{\beta} ]n \ .
\end{equation}
The parameters $\epsilon_{\alpha \beta}=\epsilon_{\alpha \beta}^*$ can be related to the quark and lepton level interactions,
\begin{eqnarray}
H_{\rm BSM}&=&\sum_{\alpha,\beta=e,\mu,\tau} \left[ {\bar \nu}_{\alpha} \gamma^{\mu}P_L \nu_{\beta} ][a^e_{\alpha \beta}{\bar e}\gamma^{\mu} e+ a^u_{\alpha \beta}{\bar u}\gamma^{\mu}u \right. \nonumber\\
 &&\hspace{3.2cm} \left. +a^d_{\alpha \beta} {\bar d}\gamma^{\mu}d+\ldots\right] \ .
\end{eqnarray}
Using ${\bar u} \gamma^{0}u = {\bar d}\gamma^{0}d = 3n$  and ${\bar e}\gamma^{0} e=n$ we have that
\begin{equation}
\label{EpsDef}
\epsilon_{\alpha \beta}= {1 \over\sqrt{2} G_F}\left(a^e_{\alpha \beta}+3a^u_{\alpha \beta}+3a^d_{\alpha \beta}\right) \ .
\end{equation}

Beyond the SM neutrino interactions can impact production, propagation and detection~\cite{Kopp:2007mi, Huber:2002bi} of neutrinos. In this paper we focus on propagation effects.

The present and next generation experiments that involve neutrino oscillations are sensitive to $\epsilon _{\alpha \beta}$'s at the percent level. 
The Super-Kamiokande I and II experiment~\cite{Mitsuka:2011ty} on atmospheric 
neutrino oscillations has put constraints that $|\epsilon_{\mu\tau}|<3.3\times10^{-2}$, $|\epsilon_{\mu\mu} - \epsilon_{\tau\tau}|<0.147$, in terms of the definition in (\ref{EpsDef}). The IceCube data are expect to improve these limits~\cite{Esmaili:2013fva}.
The proposed long baseline neutrino oscillation experiment LBNE will be sensitive to $\epsilon_{\alpha\beta}$'s in the  $0.1-0.01$ range~\cite{Adams:2013qkq}. In the more distant future a neutrino factory may be able to study matter effects that correspond to $\epsilon_{\alpha\beta}$'s at the $10^{-3}$ level.

For neutrinos interacting with light quarks, data from the precision neutrino scattering  NuTeV experiment~\cite{Davidson:2003ha} provides an important constraint. For example,  if $\varepsilon_{\mu\mu} $ arises solely from neutrino interacting with right-handed up quarks then,
$-2.4\times 10^{-2}<\varepsilon_{\mu\mu} < 0.9 \times 10^{-2}$ .

In this paper we give the contributions of all dimension six operators to the $\epsilon_{\alpha \beta}$'s under the assumption of equal number densities for electrons, protons and neutrons.  We also work out the constraints on the $\epsilon_{\alpha\beta}$'s in some  very simple extensions of the SM which contain additional scalars or gauge bosons. The models and operator analysis considered here have been discussed previously in the literature in a variety of contexts. However, we have 
explicitly expressed the $\epsilon_{\alpha\beta}$'s in terms of the coefficients of the effective operators, and presented a more complete analysis of leptoquark models.
We are focussing on $\epsilon_{\alpha \beta}$'s with magnitude great than about $10^{-3}$ and are particularly interested in knowing if in any of the models some of the $\epsilon_{\alpha \beta}$'s can be larger than the one percent level.
We show that in two models with an $SU(2)_L$ doublet leptoquark one of the diagonal elements of $\epsilon$ can be as large as $0.01-0.1$.
We discuss an interplay and complementarity between neutrino physics and leptoquark searches at hadron colliders.

Models with leptoquarks can give rise to baryon number violation from dimension four and five operators~\cite{Arnold:2013cva}. In this paper, we assume baryon number conservation.

The extensions of the SM we consider are not motivated by the hierarchy problem or intended to be correct theory of nature. Rather they are meant to play a similar role to simplified models for LHC studies. These models help us assess the most likely values of the $\epsilon_{\alpha\beta}$'s in more realistic extensions of the SM.

\section{Effective Operator Analysis}

In this section, we collect all gauge invariant effective operators made of SM fields at dimension 6, that can give non-standard interactions to neutrinos with matter. 
\begin{equation}\label{6}
H = \sum_{i=1}^{9} c_i^{\alpha\beta} O_i^{\alpha\beta} + \sum_{i=10}^{18} c_i O_i \ ,
\end{equation}
where $\alpha, \beta = e, \mu, \tau$, are lepton flavor indices, and the Wilson coefficients satisfy
$c_i^{\alpha\beta} = (c_i^{\beta\alpha})^*$
They belong to a subset of of the operators in~\cite{Grzadkowski:2010es, Buchmuller:1985jz}.  In this paper we work in a basis where the charged leptons are mass eigenstates and  the PMNS matrix arises solely from diagonalizing the neutrino mass matrix. The neutrinos are not mass eigenstates but rather are the $SU(2)_L$ partners of the mass eigenstate charged leptons, {\it i.e.}, the neutrinos are weak eigenstates.
There are three classes of such operators, those that give neutrino interactions with the charged leptons, with the light quarks and those involving the Higgs fields. 
Generally, we can write
\begin{equation}
\epsilon_{\alpha\beta} = \epsilon_{\alpha\beta}^{(l)} +  \epsilon_{\alpha\beta}^{(q)} + \epsilon_{\alpha\beta}^{(h)}  \ ,
\end{equation}
where $\epsilon^{(l,q,h)}$ are linear functions of the coefficients $c_i$.

Gauge invariance of the effective Hamiltonian in Eq.~(\ref{6}) is a powerful tool that connects new neutrino interactions to a series of ``low energy'' phenomena, such as lepton flavor violation (LFV), charged-lepton and meson decays as well as LEP data.
Therefore, the constraints obtained here are substantially stronger than those obtained in Ref.~\cite{Davidson:2003ha}, where the charged lepton interactions are only induced at loop level.
When the cutoff scale is close to the electroweak scale, higher dimensional operators that contain insertions of Higgs fields become equally important, and the connection between charged and neutral lepton interactions breaks down.
In this case, the approach of~\cite{Davidson:2003ha} is more appropriate. We will consider an example of this type in Sec.~\ref{sec:LepH}.

In this work, we are mainly interested in neutrino interactions with matter.
Therefore, for effective operators involving four leptonic fields, only those operators with at least two with electron flavor are presented.
This leaves out low energy lepton universality constraints, which require assumptions on the operators with other flavor structures not written down here. We will consider them in a specific complete model in Sec.~\ref{sec:bilepton}.

For the neutrino-quark operators we neglect the small off diagonal CKM matrix elements so that all the quark fields can be chosen to be mass eigenstate fields. 
For neutrino interactions with quarks we can also focus on operators that contain just the first generation of $u$ and $d$ quarks. This leaves out operators that have coefficients that are strongly constrained by flavor changing processes like $K^+ \rightarrow \pi^+ \nu \bar{\nu}$. We will consider such constraints when we discuss explicit models in the following section.

\subsection{Neutrino-electron Interactions}

At dimension six the gauge invariant operators relevant for non SM neutrino interactions with electrons are
\begin{eqnarray}
O_1^{\alpha\beta} &=& (\overline{L_\alpha} \gamma^\mu P_L L_\beta) (\overline{L_e} \gamma_\mu P_L L_e) \ , \\
O_2^{\alpha\beta} &=& (\overline{L_\alpha} \gamma^\mu P_L L_e) (\overline{L_e} \gamma_\mu P_L L_\beta) \ , \\
O_3^{\alpha\beta} &=& (\overline{L_\alpha} \gamma^\mu P_L L_\beta) (\bar{e} \gamma_\mu P_R e) \ ,
\end{eqnarray}
where the gauged $SU(2)_L$ indices are contracted to form a singlet within each bracket.
This is a redundant basis, the operators $O_2$ when $\alpha$ and/or $\beta$ are equal to $e$, are equivalent to $O_1$.
However, we will find it convenient to use this redundant basis.
In terms of the components of the $SU(2)_L$ doublet fields, 
\begin{eqnarray}
\label{L4}
&&H = (\overline{e_\alpha} \gamma^\mu P_L e_\beta) \left[ (c_1^{\alpha\beta} + c_2^{\alpha\beta}) (\bar e \gamma_\mu P_L e) + c_3^{\alpha\beta} (\bar e \gamma_\mu P_R e) \right] \nonumber \\
&&\ + c_1^{\alpha\beta} \left[ (\overline{\nu_\alpha} \gamma^\mu P_L \nu_\beta) (\bar e \gamma_\mu P_L e) + (\overline{\nu_e} \gamma^\mu P_L \nu_e) (\overline{e_\alpha} \gamma_\mu P_L e_\beta) \rule{0mm}{4mm}\right] \nonumber \\
&&\ + c_2^{\alpha\beta} \left[ (\overline{\nu_e} \gamma^\mu P_L \nu_\beta) (\overline{e_\alpha} \gamma_\mu P_L e) + (\overline{\nu_\alpha} \gamma^\mu P_L \nu_e) (\bar e \gamma_\mu P_L e_\beta) \rule{0mm}{4mm}\right] \nonumber \\
&&\ + c_3^{\alpha\beta} (\overline{\nu_\alpha} \gamma^\mu P_L \nu_\beta) (\bar e \gamma_\mu P_R e) + \ldots \ ,
\end{eqnarray}
where the ellipses are four neutrino operators. 

The above interactions give the following contribution to the non-standard neutrino interaction matrix defined in Eq.~(\ref{NSIdef}),
\begin{equation}
\label{epsL}
\epsilon^{(l)}_{\alpha\beta} = \frac{1}{2\sqrt{2} G_F} \left( c_1^{\alpha\beta} + c_1^{ee} \delta_e^\alpha \delta_e^\beta + c_2^{e\beta} \delta_e^\alpha + c_2^{\alpha e} \delta_e^\beta + c_3^{\alpha\beta} \right) \ .
\end{equation}

Neutrino-electron effective interactions can be constrained by the search for mono-photon events at LEP2~\cite{Berezhiani:2001rs}, similar to the dark matter search~\cite{Fox:2011fx}, in the limit when dark matter mass approaches zero. This sets a general upper limit on all elements 
\begin{equation}
\label{monogamma}
|\epsilon_{\alpha \beta}^{(l)}|<0.31 \ .
\end{equation}

The four-charged-lepton operators in the first row of Eq.~(\ref{L4}) are strongly constrained by experiment.
First, the off diagonal elements of $(c_1+c_2)$, and $c_3$ are constrained by lepton flavor violating processes~\cite{Celis:2014asa},
\begin{eqnarray}
\label{EFTLFV}
&&{\rm Br}(\mu\to3e) = \frac{|c_3^{e\mu}|^2 + 2 |c_1^{e\mu} + c_2^{e\mu}|^2}{8G_F^2} \!<\! 1.0 \times 10^{-12},  \\
&&\frac{{\rm Br}(\tau\to3e)}{{\rm Br}(\tau\to e\nu_\tau \bar \nu_e)} \!=\! \frac{|c_3^{e\tau}|^2 + 2 |c_1^{e\tau} + c_2^{e\tau}|^2}{8G_F^2} \!<\! 2.0 \times 10^{-7}, \hspace{5mm} \\
&&\frac{{\rm Br}(\tau\to\mu e^+e^-)}{{\rm Br}(\tau\to \mu\nu_\tau \bar \nu_\mu)} \!=\! \frac{|c_3^{\mu\tau}|^2 + 2 |c_1^{\mu\tau} + c_2^{\mu\tau}|^2}{8G_F^2} \!<\! 1.6 \times 10^{-7}, \hspace{5mm}
\end{eqnarray}
which typically constrain the off diagonal  $|\epsilon_{\alpha\beta}^{(l)}|$ to roughly in the range $10^{-6}$ to ${\rm a\ few}\times10^{-4}$.

The diagonal elements, on the other hand, are all constrained by the LEP2 bound on contact operators~\cite{LEP:2003aa}. 
From the $e^+e^-\to e^+e^-$ channel, we get
\begin{eqnarray}
-3.8\times10^{-3} < \frac{c_1^{ee}+c_2^{ee}}{2\sqrt{2} G_F} < 2.4\times10^{-3} \ , \label{EFTLEP}  \\
-2.3\times10^{-3} < \frac{c_3^{ee}}{2\sqrt{2} G_F} < 1.9\times10^{-3} \ .
\end{eqnarray}
From the $e^+e^-\to \mu^+\mu^-, \tau^+ \tau^-$ channels,
\begin{eqnarray}
&&-2.2\times10^{-3} < \frac{c_1^{\mu\mu}+c_2^{\mu\mu}}{2\sqrt{2} G_F}, \ \frac{c_1^{\tau\tau}+c_2^{\tau\tau}}{2\sqrt{2} G_F} <4.0\times10^{-3}, \hspace{0.5cm} \\
&&-3.7\times10^{-3} < \frac{c_3^{\mu\mu}}{2\sqrt{2} G_F} , \ \frac{c_3^{\tau\tau}}{2\sqrt{2} G_F} < 5.1\times10^{-3} \ .  \label{EFTLEP2} 
\end{eqnarray}
Therefore, typically the LEP2 constrains the diagonal magnitudes $|\epsilon_{\alpha\alpha}^{(l)}|$ to be less than ${\rm a\ few}\times 10^{-3}$.

A special case is when $c_1^{\alpha\beta} = - c_2^{\alpha\beta}$, $c_3^{\alpha\beta}=0$ for all  $\alpha$ and $\beta$. In this case, the above constraints Eqs.~(\ref{EFTLFV}--\ref{EFTLEP2}) are satisfied automatically. The flavor structure of Eq.~(\ref{L4}) immediately indicates that $\alpha\neq e$ and $\beta\neq e$, and,
\begin{equation} 
\epsilon^{(l)}_{\alpha\beta} = \frac{c_1^{\alpha\beta}}{2\sqrt{2} G_F} = - \frac{c_2^{\alpha\beta}}{2\sqrt{2} G_F} \ ,
\end{equation}
with $\alpha, \beta=\mu, \tau$.
In fact, the interactions in this case originate from a single effective operator
\begin{equation}
\label{O12}
(\overline{L_\alpha} i\sigma_2 P_R L^c_e) (\overline{L^c_\beta} i\sigma_2 P_L L_e) = \frac{1}{2} (O_1^{\alpha\beta} - O_2^{\alpha\beta}) \ .
\end{equation}

In this case, the LEP mono-photon constraint in Eq.~(\ref{monogamma}) still apply, implying all elements $|\epsilon_{\alpha \beta}|<0.31$.
We also notice the terms in the second and third lines of Eq.~(\ref{L4}) can make additional contributions to charged lepton decays.
They are $\mu\to e\nu_\mu \bar \nu_e$ and $\tau\to e\nu_\tau \bar \nu_e$, arising from the $c_2^{\mu\mu}$ and $c_2^{\tau\tau}$ terms, respectively.
Their contribution to the decay amplitudes are coherent with the SM weak interaction, thus are strongly constrained, 
With the Fermi constant determined by the electroweak observables at LEP~\cite{Marciano:1999ih},
they can be constrained by comparing the individual weak decay rate measurements with SM predictions. Setting $c_1=-c_2$ these constraints are, 
\begin{eqnarray}
\label{eq:rate1}
-1.5\times10^{-3} < \frac{c_2^{\mu\mu}}{2\sqrt{2} G_F} < 2.8\times10^{-3} \ ,\label{eq:rate1} \\
-3.9\times10^{-3} < \frac{c_2^{\tau\tau}}{2\sqrt{2} G_F} < 4.6\times10^{-3} \ .\label{eq:rate2}
\end{eqnarray}
In Eq.~(\ref{eq:rate1}) it is the error of $G_F$ that determines the range of $c_2^{\mu\mu}$, while for $c_2^{\tau\tau}$ it is the combination of the errors in $G_F$ and the rate to $\tau\to e\nu_\tau \bar \nu_e$.

There are also contributions to weak decay channels $\tau\to\mu\nu_e\bar\nu_e$, $\mu\to e\nu_\tau \bar \nu_e$ and $\tau\to e\nu_\mu \bar \nu_e$.
However, these are incoherent with the SM weak interaction amplitudes and the constraints are much weaker than the one from mono-photon.

The simplest UV complete model to obtain the operator Eq.~(\ref{O12}) from renormalizable couplings is to integrate out an $SU(2)_L$ singlet scalar with hypercharge equal to unity, that couples to a pair of lepton doublets -- a bilepton~\cite{Cuypers:1996ia}. 
In section~\ref{sec:bilepton}, we will discuss in detail this simple UV complete theory, and derive more model dependent constraints there.

\subsection{Neutrino-light-quark Interactions}
\label{EFT:lepton-quark}

The dimension six gauge invariant operators relevant for non SM neutrino interactions with quarks are
\begin{eqnarray}
O_4^{\alpha\beta} &=& (\overline{L_\alpha} \gamma^\mu P_L L_\beta) (\overline{Q_1} \gamma_\mu P_L Q_1) \ , \label{O4}\\
O_5^{\alpha\beta} &=& (\overline{L_\alpha} \gamma^\mu P_L Q_1) (\overline{Q_1} \gamma_\mu P_L L_\beta) \ , \label{O5}\\
O_6^{\alpha\beta} &=& (\overline{L_\alpha} \gamma^\mu P_L L_\beta) (\bar{u} \gamma_\mu P_R u) \ ,\label{O6} \\
O_7^{\alpha\beta} &=& (\overline{L_\alpha} \gamma^\mu P_L L_\beta) (\bar{d} \gamma_\mu P_R d) \ .\label{O7}
\end{eqnarray}
Again the gauged $SU(2)_L$ indices are contracted to form a singlet within each bracket.
In terms of the components of the $SU(2)_L$ doublet fields,
\begin{eqnarray}
\label{L2Q2}
H &=& (\overline{\nu_\alpha} \gamma^\mu P_L \nu_\beta) \left[ (c_4^{\alpha\beta} + c_5^{\alpha\beta}) (\bar u \gamma_\mu P_L u) + c_6^{\alpha\beta} (\bar u \gamma_\mu P_R u) \right. \nonumber \\
&& \hspace{2.9cm}+\left. c_4^{\alpha\beta} (\bar d \gamma_\mu P_L d) + c_7^{\alpha\beta} (\bar d \gamma_\mu P_R d) \right] \nonumber \\
&+& (\overline{e_\alpha} \gamma^\mu P_L e_\beta) \left[ (c_4^{\alpha\beta} + c_5^{\alpha\beta}) (\bar d \gamma_\mu P_L d) + c_7^{\alpha\beta} (\bar d \gamma_\mu P_R d) \right. \nonumber \\
&& \hspace{2.8cm}+\left. c_4^{\alpha\beta} (\bar u \gamma_\mu P_L u) + c_6^{\alpha\beta} (\bar u \gamma_\mu P_R u) \right] \nonumber \\
&+&\! c_5^{\alpha\beta} (\overline{\nu_\alpha} \gamma_\mu P_L e_\beta) (\bar d \gamma_\mu P_L u)\! +\! 
c_5^{\alpha\beta} (\overline{e_\alpha} \gamma_\mu P_L \nu_\beta) (\bar u \gamma_\mu P_L d).  \nonumber \\
\end{eqnarray}

The contribution of the quark operators to the $\epsilon$ matrix for neutrino non-standard interactions is,
\begin{equation}
\label{epsQ}
\epsilon^{(q)}_{\alpha\beta} = \frac{3}{2\sqrt{2} G_F} \left( 2 c_4^{\alpha\beta} + c_5^{\alpha\beta} + c_6^{\alpha\beta} + c_7^{\alpha\beta} \right) \ .
\end{equation}
Because these Wilson coefficients also appear in front of the operators in Eq.~(\ref{L2Q2}) that involve charged leptons,
the neutrino interactions are correlated with other phenomena at low energy.

First, the element $\epsilon_{e\mu}$ is connected to $\mu\to e$ conversion in the presence of a nucleus. The constraint on isospin singlet part requires~\cite{Gonzalez:2013rea}
\begin{equation}
|\epsilon_{e\mu}^{(q)}| \lesssim 10^{-7} \ .
\end{equation}

Second, the elements $\epsilon_{e\tau}$ and $\epsilon_{\mu\tau}$ may also be constrained by the LFV $\tau\to e \rho$, $\tau\to \mu \rho$ decay rates~\cite{Bergmann:1999pk}
\begin{eqnarray}
\frac{{\rm Br}(\tau\to\mu\rho^0)}{{\rm Br}(\tau\to\nu_\tau\rho^-)} = \frac{\left|c_5^{\mu\tau} +c_7^{\mu\tau}-c_6^{\mu\tau} \right|^2}{16 G_F^2} \lesssim 10^{-7} \ . \hspace{0.5cm}
\end{eqnarray}
A similar constraint holds when the muon is replaced by the electron.
Here the constraint is set on a different combination of $c_i$ from those in $\epsilon$, because $\rho$ mesons belong to an isospin triplet.
However, if only one of the effective operators in Eq.~(\ref{O4}--\ref{O7}) exists, which is the case when we 
study specific models in the next section, then this directly constrains $\epsilon$.

Third, if $c_5$ is a significant operator in the effective theory, the last row of Eq.~(\ref{L2Q2}) modifies the strength of the weak interactions.
It allows the diagonal elements of $\epsilon$ to be constrained by the lepton universality.
Namely, 
\begin{eqnarray}
\begin{split}
{\Gamma(\pi\to\mu\bar\nu_\mu) \over \Gamma(\pi\to e\bar\nu_e)} &= \frac{|2\sqrt{2} G_F+c_5^{\mu\mu}|^2}{|2\sqrt{2} G_F+c_5^{ee}|^2}, \\ 
{\Gamma(\tau\to\pi\nu_\tau) \over \Gamma(\pi\to\mu\bar\nu_\mu)} &= \frac{|2\sqrt{2} G_F+c_5^{\tau\tau}|^2}{|2\sqrt{2} G_F+c_5^{\mu\mu}|^2} \ . \\
\end{split}
\end{eqnarray}
The current limits~\cite{Pich:2013lsa, Cirigliano:2013xha, Pocanic:2014mya} then require, at leading order, at 90\% C.L.,
\begin{eqnarray}
\begin{split}
-1.6\times10^{-3} < \frac{3}{2\sqrt{2} G_F} (c_5^{\mu\mu} - c_5^{ee}) < 1.4 \times 10^{-2} \ , \hspace{0.5cm} \\  
-2.5\times 10^{-2} < \frac{3}{2\sqrt{2} G_F}(c_5^{\tau\tau}-c_5^{\mu\mu}) < 1.9 \times 10^{-3} \ . \hspace{0.5cm}
 \end{split}
\end{eqnarray}

There are also LEP2 bounds on contact operators for $e^+e^-q\bar q$ interactions~\cite{LEP:2003aa}, which only constrain the $ee$ element,
\begin{eqnarray}
&&-3.2\times10^{-2} < \frac{3(c_4^{ee}+c_5^{ee})}{2\sqrt{2} G_F}, \ \frac{3 c_4^{ee}}{2\sqrt{2} G_F} < 8.4\times10^{-2}, \hspace{0.7cm}  \\
&&-6.2\times10^{-2} < \frac{3 c_6^{ee}}{2\sqrt{2} G_F}, \ \frac{3 c_7^{ee}}{2\sqrt{2} G_F} < 4.4\times10^{-2} \label{LEPL2Q2} \ . 
\end{eqnarray}

In the next section, we study a few simple UV complete models which extend the SM by a leptoquark.
In particular, the operator in Eq.~(\ref{O5}) can be obtained by integrating out a $SU(2)_L$ singlet leptoquark at tree level,
and Eq.~(\ref{O6}) and (\ref{O7}) by integrating out a $SU(2)_L$ doublet leptoquark.
We will summarize the model dependent constraints in each model, and point out an interesting interplay between neutrino physics and leptoquark phenomenology 
at high-energy colliders.

\subsection{Operators involving the Higgs Field}

There are also dimension 6 operators with derivatives that could affect neutrino interactions. Explicitly
\begin{eqnarray}
O_8^{\alpha\beta} &=& (H^\dagger i\overleftrightarrow{D}_\mu H) (\overline{L_\alpha} \gamma^\mu P_L L_\beta) \ , \\
O_9^{\alpha\beta} &=& (H^\dagger i\overleftrightarrow{D}^a_\mu H) (\overline{L_\alpha} \tau^a \gamma^\mu P_L L_\beta) \ .
\end{eqnarray}
where $\tau^a$ are the generators acting on $SU(2)_L$ indices, and $H^\dagger\overleftrightarrow{D}_\mu H = H^\dagger D_\mu H - (D_\mu H)^\dagger H$, $H^\dagger\overleftrightarrow{D}^a_\mu H = H^\dagger \tau^a D_\mu H - (D_\mu H)^\dagger \tau^a H$.
At low energy, these operators directly modify the interaction between the $W, Z$ bosons and fermions.
They can lead to non unitarity effects in the lepton mixing PMNS matrices.
At leading order, their contribution to the matrix $\epsilon$ is
\begin{eqnarray}
\begin{split}
\epsilon_{\alpha\beta}^{(h)} &= - \frac{1}{2\sqrt{2} G_F} \left[ (c_8^{\alpha\beta}-c_9^{\alpha\beta}) 
+ 2(c_9^{\alpha e} \delta^\beta_e +c_9^{e\beta} \delta^\alpha_e)\right],
\end{split}
\end{eqnarray}
where inside the bracket the terms come from the new 
neutral ($Z\bar\nu_\alpha \nu_\beta$) and charged ($W^+ \bar \nu_\alpha e$ or $W^- \bar e \nu_\beta$) currents, respectively. 
The new charged current interaction could also affect the experimental detection of neutrinos.
We have dropped a term proportional to the unit matrix from the modified $Z\bar e e$ interaction.

The operators $O_8, O_9$ are closely related to and show up as part of the operators,
\begin{eqnarray}
(\bar L_\alpha H) i \!\!\not\! \partial (H^\dagger L_\beta), \ \ (\bar L_\alpha \tau^a H) i \!\!\not\! \partial (H^\dagger \tau^a L_\beta) \ ,
\end{eqnarray}
which can be obtained by integrating out $SU(2)_L$ singlet or triplet heavy fermions in the Seesaw mechanisms~\cite{seesaw, Foot:1988aq} for generating Majorana neutrino masses. 
Their effects on the matrix $\epsilon$ have already been tightly constrained to be less than $\lesssim10^{-3}$~\cite{Antusch:2008tz}. We will not consider complete models for these operators in this paper. However,
we would like to point out that these kinetic operators may have an interesting connection to cosmology.
During the electroweak phase transition, the Higgs field has non-vanishing time derivative. These operators result in chemical potential terms for lepton number,
$ \sim \partial_\mu (H^\dagger H) (\bar L_\alpha \gamma^\mu L_\beta)$.
In the presence of sphaleron and/or other $B$ and $L$ violating processes these kinetic operators could provide an explanation for the baryon asymmetry in the Universe. This would require a strong first electroweak phase transition in order to obtain a sufficient large time derivative and baryon number.

For completeness, we also list the operators that do not directly involve neutrino fields (see Table 2 of~\cite{Grzadkowski:2010es}),
\begin{eqnarray}
O_{10} &=& (H^\dagger i\overleftrightarrow{D}_\mu H) (\bar{e} \gamma^\mu P_R e) \ , \nonumber \\
O_{11} &=& (H^\dagger i \overleftrightarrow{D}_\mu H) (\overline{Q_1} \gamma^\mu P_L Q_1) \ , \nonumber \\
O_{12} &=& (H^\dagger i \overleftrightarrow{D}^a_\mu H) (\overline{Q_1} \tau^a \gamma^\mu P_L Q_1) \ , \nonumber \\
O_{13} &=& (H^\dagger i\overleftrightarrow{D}_\mu H) (\bar{u} \gamma^\mu P_R u) \ , \nonumber \\
O_{14} &=& (H^\dagger i\overleftrightarrow{D}_\mu H) (\bar{d} \gamma^\mu P_R d) \ , \nonumber \\
O_{15} &=& (H^\dagger D^\mu H)^* (H^\dagger D_\mu H) \ , \nonumber \\
O_{16} &=& (H^\dagger H) W^a_{\mu\nu} W^{a\mu\nu} \ , \nonumber \\
O_{17} &=& (H^\dagger H) B_{\mu\nu} B^{\mu\nu} \ , \nonumber \\
O_{18} &=& (H^\dagger \tau^a H) W^a_{\mu\nu} B^{\mu\nu} \ . \nonumber
\end{eqnarray}
They can modify the neutral current interactions mediated by the $Z$ boson on the source side of the matter potential, thus the contributions to $\epsilon$ is proportional to a unit matrix.
As a result, these effects will not be measured by neutrino oscillation experiments, but can be tested in the other precision measurements.

\subsection{$d=8$ operators}

At dimension 8  the connection between charged lepton and neutrino interactions can be broken~\cite{Berezhiani:2001rs, Friedland:2011za, Davidson:2011kr} by inserting Higgs fields. Hence the constraints on neutrino matter interactions from dimension 8 operators are weaker. However, it seems difficult to get dimension eight operators without generating at leas some of the dimension six operators~\cite{Antusch:2008tz}.

\section{Simple Renormalizable Models}

In this section, we will go through several simple extensions of the SM that have non standard neutrino interactions.
They can be viewed as UV completions of the effective Hamiltonian discussed above.
As we will show, in addition to the generic bounds derived in the effective theory,
often there are additional model dependent constraints.
A survey of some simple models allows us to get a better and more realistic sense of of the  values  the elements of the matrix $\epsilon$ will take in extensions of the SM.

In all the simple models we consider that involve an additional scalar representation of the gauge group, the $\epsilon _{\alpha \beta}$'s are not all independent but rather satisfy the relations,
\begin{equation}
\label{relationepsilons}
|\epsilon_{e \mu}|^2=\epsilon_{ee}\epsilon_{\mu\mu},~~|\epsilon_{\mu \tau}|^2=\epsilon_{\mu\mu}\epsilon_{\tau\tau},~~|\epsilon_{e \tau}|^2=\epsilon_{ee}\epsilon_{\tau\tau} \ .
\end{equation}
For some of the models, their impact on neutrino oscillations have been studied previously in the literature. We include them for completeness.
For the leptoquark models, we present a more complete analysis of their impact on neutrino physics than was done in the previous literature.

\subsection{$SU(2)_L$ Singlet Bilepton}
\label{sec:bilepton}

The simplest model with non-standard neutrino interactions has an additional  $SU(2)_L$ singlet scalar $S$, which couples to lepton doublets,
\begin{equation}
\mathcal{L} = \lambda_{\alpha\beta} \overline{L^c_\alpha} (i\sigma_2) P_L L_\beta S + {\rm h.c.} \ .
\end{equation}
Here $S$ has electric charge $+1$, and $P_L=(1-\gamma_5)/2$ is the left handed projection operator. The indices $\alpha, \beta$ must be antisymmetric, $\lambda_{\alpha\beta}=-\lambda_{\beta\alpha}$. 
In general there are only three independent complex couplings, and the above Lagrangian can be decomposed in the flavor space as
\begin{eqnarray}
\label{lagrangian1}
\mathcal{L} &=& 2\lambda_{e\mu} (\bar \nu_e^c P_L \mu - \bar \nu_\mu^c P_L e) S + 2\lambda_{\mu\tau} (\bar \nu_\mu^c P_L \tau - \bar \nu_\tau^c P_L \mu) S \nonumber \\
&&+ 2\lambda_{\tau e} (\bar \nu_\tau^c P_L e - \bar \nu_e^c P_L \tau) S + {\rm h.c.} \ .
\end{eqnarray}
Then the couplings of relevance to the matter effects in neutrino oscillations are $\lambda_{e\mu}$ and $\lambda_{\tau e}$.
Integrating out the bilepton $S$ and matching on to the operator analysis, we find that,
\begin{eqnarray}
c_1^{\alpha\beta} = - c_2^{\alpha\beta} = \frac{2\lambda_{e\alpha} \lambda_{e\beta}^*}{m_S^2} \ ,
\end{eqnarray}
and the other $c_i=0$, which implies
\begin{eqnarray}
\epsilon_{\alpha\beta} = \frac{\lambda_{e\alpha} \lambda_{e\beta}^*}{\sqrt{2} G_F m_S^2}, \ \ \ (\alpha, \beta = \mu, \tau) \ ,
\end{eqnarray}
and $\varepsilon_{ee}=\varepsilon_{e\mu}=\varepsilon_{e\tau}=0$, and the relation Eq.~(\ref{relationepsilons}) holds. Note that in this model $\epsilon_{\mu\mu}$ and $\epsilon_{\tau \tau}$ are positive but $\epsilon_{\mu \tau}$ can be complex.

As mentioned in the previous section, because of the property $c_1^{\alpha\beta} = - c_2^{\alpha\beta}$, $c_3^{\alpha\beta}=0$, and therefore, none of the model
independent constraints in Eqs.~(\ref{EFTLFV}--\ref{EFTLEP2}) apply. 

There are important constraints on the Yukawa couplings of $S$ from experimental limits on charged lepton flavor violation.
\begin{eqnarray}
{\rm Br}(\mu\to e \gamma) = \frac{\alpha}{48\pi} \frac{1}{G_F^2 m_S^4} |\lambda_{\tau e} \lambda_{\mu\tau}|^2 &<& 1.2\times 10^{-11} \ , \hspace{0.5cm} \\
{\rm Br}(\tau\to e \gamma) = \frac{\alpha}{48\pi} \frac{1}{G_F^2 m_S^4} |\lambda_{e\mu} \lambda_{\mu\tau}|^2 &<& 3.3\times 10^{-8} \ , \\
{\rm Br}(\tau\to \mu \gamma) = \frac{\alpha}{48\pi} \frac{1}{G_F^2 m_S^4} |\lambda_{e\mu} \lambda_{\tau e}|^2 &=&  \frac{\alpha}{24\pi} |\epsilon_{\mu\tau}|^2  \nonumber \\
&<& 4.4\times 10^{-8} \ .
\end{eqnarray}
The limit on the branching ratio for the charge lepton flavor violating decay $\tau \rightarrow \mu \gamma$ gives the constraint, $|\epsilon_{\mu\tau}|<0.021$.

The experimental limit on the branching ratio muon radiative decay is quite strong. However, the rate for $\mu\to e \gamma$ vanishes as $\lambda_{\mu \tau} \rightarrow 0$ and in this does not restrict the values of the neutrino matter interaction parameters.  When  $\lambda_{\mu \tau}=0$ the Yukawa couplings of $S$ have the continuos  global symmetry: $
(L_e, e_R)  \rightarrow  e^{-2i \alpha} (L_e, e_R),~ (L_\mu, \mu_R)  \rightarrow  e^{i \alpha} (L_\mu, \mu_R),~(L_\tau, \tau_R)  \rightarrow  e^{i \alpha} (L_\tau, \tau_R),~ S\rightarrow  e^{i \alpha}S $ . This global symmetry cannot be exact but it is only broken by very small neutrino mass terms. The presence of this symmetry ensures that even if the Yukwa couplings of $S$ are large radiative corrections will not induce a significant value for $\lambda_{\mu \tau}$. Henceforth we neglect the coupling $\lambda_{\mu \tau}$.

Integrating out $S$ at tree level gives a new contribution to the effective Hamiltonian for the weak decays of charged leptons,
\begin{equation}
\begin{split}
H_{\rm eff} &= 2\sqrt{2}G_F \epsilon_{\mu\mu} ({\bar e_L}\gamma^{\alpha}P_L \nu_e)({\bar \nu}_\mu \gamma_{\alpha} P_L \mu) \\
&+ 2\sqrt{2}G_F \epsilon_{\tau\tau} ({\bar e_L}\gamma^{\alpha}P_L \nu_e)({\bar \nu}_\tau \gamma_{\alpha} P_L \tau) + {\rm h.c.} \ ,
\end{split}
\end{equation}
where the correction to $\tau\to\mu\nu_\tau\bar \nu_\mu$ decay has been suppressed due to the assumed smallness of $\lambda_{\mu \tau}$.

The most important constraints arise from lepton universality. For example, the ratio of the weak decay rates is related to the epsilons,
\begin{eqnarray}
\frac{\Gamma(\tau\to\mu\nu_\tau\bar \nu_\mu)}{\Gamma(\tau\to e\nu_\tau\bar \nu_e)} = \frac{1}{(1+\epsilon_{\tau\tau})^2} \ , 
\end{eqnarray}
The experimental constraint on such ratio (see the table 2 in Ref.~\cite{Pich:2013lsa}) requires, at 90\% C.L. (1.65\,$\sigma$), 
\begin{eqnarray}
\label{Eq:bilepWeakDecaymu}
 \epsilon_{\tau\tau} < 2.5 \times 10^{-4} \ .
\end{eqnarray}
Similarly, experimental constraints on the other ratios ${\Gamma(\tau\to e\nu_\tau\bar \nu_e)}/{\Gamma(\mu\to e\nu_\mu\bar \nu_e)}, {\Gamma(\tau\to\mu\nu_\tau\bar \nu_\mu)}/{\Gamma(\mu\to e\nu_\mu\bar \nu_e)}$ requires, at 90\% C.L.,
\begin{eqnarray}\label{Eq:bilepWeakDecaytau}
\begin{split}
&2.5\times10^{-3} < \epsilon_{\tau\tau} - \epsilon_{\mu\mu} < 6.8 \times 10^{-4}, \\ 
&\hspace{2cm}\epsilon_{\mu\mu} \ll 10^{-4} \ .
\end{split}
\end{eqnarray}
Here the experimental significance is high for the second ratio to be positive, leaving very little room for $\epsilon_{\mu\mu}$ to contribute. 
We notice the epsilons can also be constrained with the individual decay rate, as discussed in Eq.~(\ref{eq:rate1}) and (\ref{eq:rate2}). 
We find they give weaker constraints on the epsilons than above.

Using the limits in Eqs.~(\ref{Eq:bilepWeakDecaymu}) and (\ref{Eq:bilepWeakDecaytau}) the relation in Eq.~(\ref{relationepsilons}) implies that
$|\epsilon_{\mu\tau}|$ is also tiny, $\ll 10^{-4}$. 
Note that this is stronger than the limit from $\tau \rightarrow \mu \gamma$.

Finally we note that there is a new one-loop contribute to the anomalous magnetic moment of the muon (setting $\lambda_{\mu\tau}=0$).
\begin{equation}
\delta \left(\frac{g-2}{2}\right)_\mu = - \frac{m^2_\mu}{12\pi^2 m_S^2}  |\lambda_{e\mu}|^2 =-1.6\times 10^{-9} \epsilon_{\mu \mu} \ .
\end{equation}
Experimentally, $(g-2)_\mu/2=(11 659 208.0\pm6.3)\times 10^{10}$~\cite{Bennett:2006fi}. There is a $3\sigma$ deviation from the SM prediction.
This new contribution is in the opposite direction from the observed discrepancy. In any case, the muon decay constraint in Eq.~(\ref{Eq:bilepWeakDecaytau}) means it is too small to impact measurements of the anomalous magnetic moment of the muon.

To summarize, in the bilepton model, the only non-vanishing elements of epsilon are $\epsilon_{\mu\mu}$, $\epsilon_{\mu\tau}$ and $\epsilon_{\tau\tau}$, and the present experimental limits already constrain them to be no larger than ${\rm a\ few}\times10^{-4}$.

\subsection{Leptophilic $SU(2)_L$ Doublet Scalar}
\label{sec:LepH}

The second model we consider contains a scalar doublet $S = (H^+, H^0)^T$, carrying the same quantum numbers as the SM Higgs doublet. Here we assume it has no VEV, so all the components are physical. Its Yukawa couplings with leptons take the form
\begin{equation}
\begin{split}
\mathcal{L} &= \lambda_{ij} \overline{L_i} P_R e_j S + {\rm h.c.} \\
&= \lambda_{ij} (\overline{\nu_i} P_R e_j H^+ + \overline{e_i} P_R e_j H^0) + {\rm h.c.} \ .
\end{split}
\end{equation}
Direct search at the LEP2 experiment constrains the charged scalar $H^+$ to be heavier than 103\,GeV.
If $H^+$ mainly decays into a electron (or muon) and a neutrino, the LHC constraint on slepton NLSP~\cite{CMSslepton} can be applied to $H^+$, which requires it to be heavier than 290\,GeV.
Moreover, precision electroweak physics implies that $|m_{H^0}-m_{H^+}|\lesssim107\,{\rm GeV}$ at 2\,$\sigma$~\cite{Barbieri:2006dq}.  
Therefore, if we split the charged and neutral components

We first neglect the mass difference between the charged and neutral components of $S$. Integrating out $S$ and matching on to the operator analysis yields,
\begin{eqnarray}
c_3^{\alpha\beta}=\frac{2\lambda_{\alpha e} \lambda_{\beta e}^*}{m_S^2} \ ,
\end{eqnarray}
and the other $c_i=0$, which implies 
\begin{eqnarray}
\epsilon_{\alpha\beta} = -\frac{\lambda_{\alpha e} \lambda_{\beta e}^*}{4\sqrt{2} G_F m_S^2}, \ \ \ (\alpha, \beta = e, \mu, \tau) \ .
\end{eqnarray}
In this model all the diagonal elements $\epsilon_{\alpha\alpha}$ are negative.
Here the model
independent constraints Eqs.~(\ref{EFTLFV}--\ref{EFTLEP2}) apply. 
All the off diagonal epsilon elements constrained to be less than $10^{-4}$.  
The direct constraints on the diagonal elements allow them to be as as large as ${\rm a\ few}\times10^{-3}$, but in this model Eq.~(\ref{relationepsilons}) further forces all but one to be less than $10^{-4}$.

However, if the mass scale $m_S$ is not far above the electroweak scale, which is allowed by the LEP2 constraints, the mass splitting between $H^+$ and $H^0$ could have a significant effect. Namely, the cutoff scale in the effective theory language is no longer gauge invariant. In this general case, the effective Hamiltonian takes the form
\begin{eqnarray}
\begin{split}
H &= c_3^{\alpha\beta} \left[ (\overline{\nu_\alpha} \gamma^\mu P_L \nu_\beta) (\bar e \gamma_\mu P_R e) \rule{0mm}{5mm}\right. \\
& \hspace{1.2cm}+\left. \frac{m_{H^+}^2}{m_{H^0}^2}(\overline{e_\alpha} \gamma^\mu P_L e_\beta) (\bar e \gamma_\mu P_R e) \rule{0mm}{5mm}\right] \ ,
\end{split}
\end{eqnarray}
where $c_3^{\alpha\beta}={2\lambda_{\alpha e} \lambda_{\beta e}^*}/{m_{H^+}^2}$.
Therefore, if ${m_{H^+}}<{m_{H^0}}$, all the constraints on $c_3^{\alpha\beta}$ in Eqs.~(\ref{EFTLFV}--\ref{EFTLEP2}) can be relaxed by a factor of $({m_{H^0}}/{m_{H^+}})^2$. In the most optimistic case, choosing $m_{H^+}$ equal to the current collider limits and taking account of and relations Eq.~(\ref{relationepsilons}), we find one of the diagonal elements of the $\epsilon$ matrix can be as larger as one percent level,
\begin{eqnarray}
 &\epsilon_{\tau\tau} < 1.5\times 10^{-2},& \nonumber \\
 &\epsilon_{ee} < 0.43\times 10^{-2}, \ \epsilon_{\mu\mu} < 0.69\times 10^{-2}\ .& 
\end{eqnarray}
We note this case is accompanied with the prediction of a light ($\sim 100\,$GeV) charged scalar $H^+$ decaying into a charged lepton and neutrino, which could be probed with future colliders.

\subsection{$SU(2)_L$ Singlet Leptoquark}

The simplest leptoquark model that gives neutrino non-standard interaction is
\begin{equation}
\begin{split}
\label{LQ2}
\mathcal{L} &= \lambda_{ij} \overline{L^c_{i}} (i\sigma_2) P_L Q_{j} S + \lambda'_{ij} \overline{u_{i}^c} P_R e_{j} S + {\rm h.c.} \\
&= \lambda_{ij} (\overline{\nu^c_{i}} P_L d_{j} - \overline{e^c_{i}} P_L u_{j}) S + \lambda'_{ij} \overline{u_{i}^c} P_R e_{j} S + {\rm h.c.}, 
\end{split}
\end{equation}
where $S$ is a $SU(2)$ singlet leptoquark with hypercharge $2/3$.

Since we are interested in neutrino interactions in the flavor basis, we choose to work in the basis where the down type quark  mass matrices are diagonal.
In the language of effective operators discussed in Eq.~(\ref{L2Q2}), the singlet leptoquark model gives 
\begin{eqnarray}
c_4^{\alpha\beta} = -c_5^{\alpha\beta} = \frac{\lambda_{\alpha 1} \lambda_{\beta 1}^*}{2 m_S^2} \ ,
\end{eqnarray} 
for all the $\alpha, \beta = e, \mu, \tau$ and the other $c_i=0$.. 
The non-standard neutrino matter interaction parameters are related to the Yukawa couplings by,
\begin{eqnarray}
\epsilon_{\alpha\beta} = {3 \over 4} \frac{\lambda_{\alpha 1} \lambda_{\beta 1}^*}{\sqrt{2} G_F m_S^2}, \ \ \ (\alpha, \beta =e, \mu, \tau) \ ,
\end{eqnarray}
The flavor diagonal elements $\epsilon_{ee}$, $\epsilon_{\mu\mu}$ and $\epsilon_{\tau\tau}$ are real and positive, while the flavor-changing ones are complex in general. 

The model independent constraints in sec.~\ref{EFT:lepton-quark} apply here. Again the key point is
$SU(2)_L$ gauge invariance relates neutrino interactions with quarks to those of the charged leptons.
We summarize these constraints here,
\begin{eqnarray}
\begin{split}
&|\epsilon_{e \mu}|<10^{-7}, \ 
|\epsilon_{\mu\tau}| < 9.2 \times 10^{-4}, \
|\epsilon_{e\tau}| < 1.1 \times 10^{-3}, \hspace{0.7cm} \\
&\hspace{1cm}-1.6\times10^{-3} < \epsilon_{\mu\mu} - \epsilon_{ee} < 1.4 \times 10^{-2} \ , \\  
&\hspace{1cm}-2.5\times 10^{-2} < \epsilon_{\tau\tau}-\epsilon_{\mu\mu} < 1.9 \times 10^{-3} \ . 
\end{split}
\end{eqnarray}
where arise from LFV decays and lepton universality.

The LEP2 contact operator bounds for $e_L^+e_L^-\to u_L\bar u_L$,  require $|\epsilon_{ee}|<8.4\times10^{-2}$.
The LHC search for leptoquark pair production puts a constraint on the mass of $S$ to be larger than 780\,GeV~\cite{CMSLQ}. 
If the couplings $\lambda$ are equal to unity, this implies all the $\epsilon_{\alpha\beta}$'s can be most a few percent.

Combining these limits with the relations in Eq.~ (\ref{relationepsilons}), we can also get constraints on each of the diagonal elements
\begin{eqnarray}
\epsilon_{ee}, \ \epsilon_{\mu\mu} < 2.6 \times 10^{-3}, \ \ \ \epsilon_{\tau\tau} < 4.5 \times 10^{-3} \ . 
\end{eqnarray}

Now, we turn to  the constraints from experimental results on flavor changing effects in the quark sector.
Eq.~(\ref{LQ2}) can be rewritten in terms of the mass eigenstate quarks fields and the CKM matrix elements,
\begin{eqnarray}
\mathcal{L} &=& - (\lambda_{\alpha 1} - \lambda_{\alpha 2} \sin\theta_C) \overline{e^c_\alpha} P_L u S - \lambda_{\alpha 1} \overline{\nu^c_\alpha} P_L d S \nonumber \\
& & - (\lambda_{\alpha 1} \sin\theta_C + \lambda_{\alpha2}) \overline{e^c_\alpha} P_L c S  -\lambda_{\alpha 2} \overline{\nu^c_\alpha} P_L s S \nonumber \\
& & + \ldots +{\rm h.c.} \ ,
\end{eqnarray}
where $\theta_C$ is the Cabibbo angle. A very strong bound on $\lambda_{\alpha2}$ comes from the measured branching ratio ${\rm Br}(K^+ \to \pi^+ \nu {\bar \nu})=(1.7\pm 1.1)\times 10^{-10}$. The contribution to this decay rate from leptoquark exchange alone  is (there is an interference piece with the SM contribution we have neglected), 
\begin{equation}
\Gamma (K^+\to \pi^+\nu\bar \nu) = \frac{m_K^5}{24576 \pi^3 m_S^4} ( \lambda^\dagger \lambda)_{11} ( \lambda^\dagger \lambda)_{22} \ .
\end{equation}
Demanding that this contribution is less than the measured branching ratio requires,
\begin{equation}
\left( {3 \over 4}{ (\lambda^\dagger \lambda)_{11} \over \sqrt{2} G_F m_S^2} \right)\left( {3 \over 4}{ (\lambda^\dagger \lambda)_{22} \over \sqrt{2} G_F m_S^2} \right)< 4.9\times 10^{-10} \ .
\end{equation}
The first bracket is nothing but the trace of the epsilon matrix $(\epsilon_{ee}+\epsilon_{\mu \mu}+\epsilon_{\tau \tau})$.
For neutrino experimental prospects, we are focussing on the case where at least one of the  $\epsilon_{\alpha\beta}$ is of order $10^{-3}$ and because of the relationship between the off diagonal and diagonal epsilons in this model that implies that one of the diagonal $\epsilon_{\alpha \alpha}$ is of order $10^{-3}$. Since all the $\epsilon_{\alpha\alpha}$'s (for $\alpha=e, \mu, \tau$) are positive, the above constraint implies that
$(\lambda^\dagger \lambda)_{22} << 10^{-3} (\lambda^\dagger \lambda)_{11}$.

Next, we consider the bound coming from the $D-\bar D$ mixing. Integrating out $S$ in the box diagram yields the $\Delta c=2$ interaction,
\begin{eqnarray}
H_{\Delta c=2} 
&=& \frac{[(\lambda^\dagger \lambda)_{11}  \sin\theta_C ]^2}{128\pi^2 m_S^2} (\bar c_L \gamma^\mu u_L)(\bar c_L \gamma_\mu u_L) \ .
\end{eqnarray}
where we have neglected the terms involving $\lambda_{\alpha2}$ in light of the above $K^+\to \pi^+\nu\bar \nu$ decay bound.
Using this Hamiltonian, we obtain the relation between the $\epsilon_{\alpha\beta}$'s and the new contribution to $D$ meson mass difference,
\begin{equation}
\delta(m_{D_1^0} - m_{D_2^0})
= \frac{G_F^2 m_S^2 f_D^2 m_D}{54\pi^2} \sin^2\theta_C |\epsilon_{ee}+\epsilon_{\mu\mu}+\epsilon_{\tau\tau}|^2 \ .
\end{equation}
We require this new contribution does not exceed the experimentally measured mass difference, which implies
\begin{equation}\label{LQDDbar}
\epsilon_{ee}+\epsilon_{\mu\mu}+\epsilon_{\tau\tau}< 3.1\times 10^{-3} \left( \frac{1\,\rm TeV}{m_S} \right) \ .
\end{equation}
There are several interesting implications from this bound. First, because all diagonal epsilons in this model are the positive, Eq.~(\ref{LQDDbar}) also sets the upper bound on the individual $\epsilon_{\alpha\alpha}$. Second, as the future collider limit pushes the leptoquark mass to higher scale, it improves the bound on the diagonal elements of the matrix $\epsilon$ at the same time.

To summarize, we surveyed the present experimental constraints from the present LHC and low energy experiments, and found in the singlet leptoquark model, all the epsilon elements are  constrained to be less than ${\rm a\ few}\times10^{-3}$.

\subsection{$SU(2)_L$ Doublet Leptoquark}

Next we consider a model with a scalar leptoquark that is a doublet under $SU(2)_L$ and has hypercharge $-7/3$. The Yukawa couplings of $S$ to the quarks and leptons are given by,
\begin{equation}
\begin{split}
\label{lagrangeleptodoublet}
\mathcal{L} &= \lambda_{ij} \overline{L_i} P_R u_j S + \lambda'_{ij} \overline{Q_i} P_R e_j \tilde S + {\rm h.c.} \\
&= \lambda_{ij} (\overline{\nu_i} P_R u_j X + \overline{e_i} P_R u_j Y) \\ 
& + \lambda'_{ij} (\overline{u_i} P_R e_j Y - \overline{d_i} P_R e_j X) + {\rm h.c.} \ .
\end{split}
\end{equation}
In terms of their components, $S$ and $\tilde S$ are,
\begin{eqnarray}
S = \begin{pmatrix}
X \\
Y
\end{pmatrix}, \ \ \ \tilde S = \begin{pmatrix}
Y \\
-X
\end{pmatrix} \ .
\end{eqnarray}
Precision electroweak physics implies that $|m_X-m_Y|\lesssim62\,{\rm GeV}$~\cite{Keith:1997fv}.  Therefore we neglect the splitting between the $X$ and $Y$ scalars setting $m_X=m_Y=m_S$.

In Eq.~(\ref{lagrangeleptodoublet}) we choose to be in the basis where the up type quark and charged lepton mass matrices are diagonal. 
In the effective language, integrating out $S$ yields 
\begin{eqnarray}
c_6^{\alpha\beta} = -  \frac{\lambda_{\alpha 1} \lambda_{\beta 1}^*}{2 m_S^2} \ ,
\end{eqnarray} 
with the other $c_i=0$.
Then the neutrino matter interaction parameters are related to the Yukawa couplings by,
\begin{eqnarray}
\label{EpsLQ}
\epsilon_{\alpha\beta} = -{3 \over 4} \frac{\lambda_{\alpha 1} \lambda_{\beta 1}^*}{\sqrt{2} G_F m_S^2}, \ \ \ (\alpha, \beta =e, \mu, \tau) \ ,
\end{eqnarray}
where the elements $\epsilon_{ee}$, $\epsilon_{\mu\mu}$ and $\epsilon_{\tau\tau}$ are real and negative, while the flavor-changing ones are complex in general.

Like the singlet leptoquark case, the $SU(2)_L$ invariance again implies the generic LFV decay constraints apply, and we have
\begin{equation}
|\epsilon_{e \mu}|<10^{-7}, \ |\epsilon_{\mu\tau}| < 9.2 \times 10^{-4}, \ |\epsilon_{e\tau}| < 1.1 \times 10^{-3} \ .
\end{equation}
However, the lepton universality constraints no longer apply because the new contributions have negligible interference with the SM weak decay amplitudes.
The element $\epsilon_{ee}$ can also be constraint by LEP2 contact operators bound for $e_L^+e_L^-\to u_R\bar u_R$, which requires $|\epsilon_{ee}|<6.2\times10^{-2}$.

LHC data constraints on leptoquark pair production. If the leptoquark  couples only to the light quarks and charged leptons then $m_Y\gtrsim1\,{\rm TeV}$~\cite{CMSLQ}. 
Here the LHC constraint is stronger compared singlet leptoquark case, because the branching ratio to a charged lepton and a jet is larger.
For couplings $\lambda$ equal to unity,  LHC data constrains the magnitude of the $\epsilon_{\alpha \beta}$'s to be at most  a few percent.

Taking account of the relations among the  in Eq.~(\ref{relationepsilons}), we find
a simple way to satisfy all of these constraints is to have only one of $\epsilon_{ee}$, $\epsilon_{\mu \mu}$ and $\epsilon_{\tau \tau}$ be sizable. 
In terms of the Yukawa couplings $\lambda$ that can occur if all them are very small except one of $\lambda_{e1}$, $\lambda_{\mu1}$  and $\lambda_{\tau 1}$.

There are also constraints from say $D-\bar{D}$ mixing on $(\lambda^\dagger \lambda)_{21}$. They are satisfied if all elements of $\lambda$ other than the first column are negligibly small.
The couplings $\lambda'$ are constrained by flavor changing processes like $K -{\bar K}$ mixing and since they do not impact neutrino physics we also take them to be very small. 

Given our ignorance of the origin of flavor it is conceivable that such relations could hold at some higher scale. However the various couplings mix under renormalization and so even if we impose these constraints at a high scale we should check they are still satisfied at low energies.
The renormalization group flow of the couplings is restricted by the transformation properties of the Yukawa couplings under the flavor group $G_F= SU(3)_Q \times SU(3)_u \times SU(3)_d \times SU(3)_L  \times SU(3)_e$.   The representations of coupling constant spurions are

\begin{eqnarray}
\begin{tabular}{cccccc}
\hline
&  $SU(3)_Q$ & $SU(3)_u$ & $SU(3)_d$ & $SU(3)_L$ & $SU(3)_e$ \\
\hline
\vspace{-0.3cm}\\
$Y_u$ & 3 & $\bar 3$ & 1 & 1 & 1 \\
$Y_d$ & 3 & 1 & $\bar 3$ & 1 & 1 \\
$Y_e$ & 1 & 1 & 1 & 3 & $\bar 3$ \\
$\lambda$ & 1 & $\bar 3$ & 1 & 3 & 1 \\
$\lambda'$ & 3 & 1 & 1 & 1 & $\bar 3$\\
\hline
\end{tabular} \nonumber
\end{eqnarray}

where we define the Standard Model Yukawa couplings as
\begin{eqnarray}
\begin{split}
\mathcal{L}_Y &= (Y_u)_{ij} \overline{Q_i} (i\tau_2) H^* P_R u_j  + (Y_d)_{ij} \overline{Q_i} H P_R d_j \\
&+ (Y_e)_{ij} \overline{L_i} H P_R e_j X + {\rm h.c.} \ .
\end{split}
\end{eqnarray}
The one-loop beta functions are
\begin{eqnarray}
16\pi^2 \frac{d Y_u}{d\ln\mu} &=& \frac{3}{2} (Y_u Y_u^\dagger - Y_d Y_d^\dagger) Y_u + 2 \lambda' Y_e^\dagger \lambda \nonumber \\
&&+ \frac{1}{2} \lambda' \lambda'^\dagger Y_u + Y_u \lambda^\dagger \lambda + \ldots \ ,  \\
16\pi^2 \frac{d Y_d}{d\ln\mu} &=& - \frac{3}{2} (Y_u Y_u^\dagger - Y_d Y_d^\dagger) Y_d \nonumber \\
&&+ \frac{1}{2} \lambda' \lambda'^\dagger Y_d + \ldots \ ,  \\
16\pi^2 \frac{d \lambda}{d\ln\mu} &=& -2 Y_e \lambda'^\dagger Y_u + \lambda Y_u^\dagger Y_u + \frac{1}{2} Y_e Y_e^\dagger \lambda \nonumber \\
&&+ \frac{5}{2} \lambda \lambda^\dagger \lambda + \ldots \ ,  \\
16\pi^2 \frac{d \lambda'}{d\ln\mu} &=& -2 Y_u \lambda^\dagger Y_e + \frac{1}{2} (Y_u Y_u^\dagger + Y_d Y_d^\dagger) \lambda'  \nonumber \\
&& + \lambda' Y_e^\dagger Y_e + \frac{7}{2} \lambda' \lambda'^\dagger \lambda' + \ldots \ .
\end{eqnarray}
where the ellipses represents those radiative corrections proportional to gauge couplings or the trace of Yukawa matrices, {\it i.e.}, those do not modify the flavor structure of the matrices.

For illustration, we take an ansatz for the structure of the Yukawa matrices at the cutoff scale $\Lambda$ discussed previously,
\begin{eqnarray}
\label{ansatz}
Y_u(\Lambda) &=& \frac{\sqrt{2}}{v} \begin{pmatrix}
m_u &0 &0 \\
0& m_c &0 \\
0&0 & m_t 
\end{pmatrix}, \ \ \ \lambda(\Lambda) = \begin{pmatrix}
\lambda_{e1} & 0&0 \\
\lambda_{\mu1}& 0 & 0\\
\lambda_{\tau1}&0 &0 
\end{pmatrix}, \nonumber \\
Y_d(\Lambda) &=& \frac{\sqrt{2}}{v} V \begin{pmatrix}
m_d & 0&0 \\
0& m_s &0 \\
0&0 & m_b 
\end{pmatrix} V^\dagger, \ \ \ \lambda'(\Lambda)=0 \ ,
\end{eqnarray}
where $V$ is the CKM matrix, and only one of the couplings $\lambda_{e1}$, $\lambda_{\mu1}$, $\lambda_{\tau 1}$, is non-zero.

\begin{figure*}
\includegraphics[width=2.10\columnwidth]{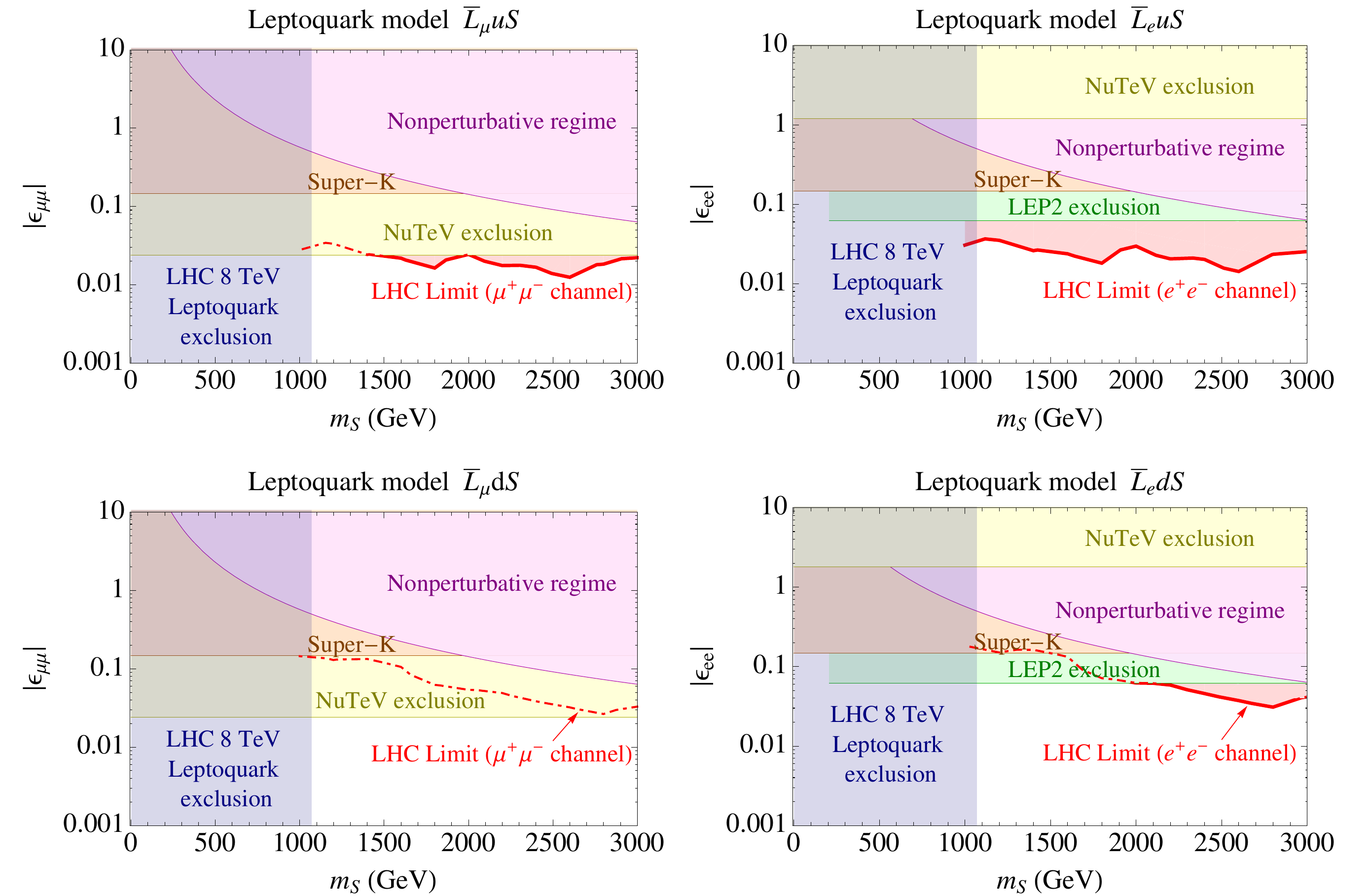}
\caption{Red curves: upper bounds on $|\epsilon_{\mu\mu}|$ (left) and $|\epsilon_{ee}|$ (right) from LHC data on $\mu^+\mu^-$, $e^+e^-$ pair production rates, in $\bar L u S$ (first row) and $\bar L d S$ (second row) leptoquark models.
The blue region is the LHC exclusion of the mass of leptoquark from its pair production and decays.
The yellow region is excluded by NuTeV experiment on neutrino-quark scatterings.
The orange region is excluded by the Super-K experiment as mentioned in the introduction.
The LEP2 exclusion on $|\epsilon_{ee}|$ from $e^+e^-\to u \bar u$ channel is the green region.
The magenta region is non-perturbative regime when the leptoquark coupling $|\lambda_{\alpha i}|>\sqrt{4\pi}, \ (\alpha=e, \mu, \ i=u,d)$.
All shaded regions are excluded.} \label{LHC}
\end{figure*}

Our ansatz for the structure of the $S$ Yukawa couplings at the scale $\Lambda$ in Eq.~(\ref{ansatz}) is not consistent with a diagonal up quark mass matrix and the couplings $\lambda'$ being zero.
 At a low scale $\mu$, the up quark Yukawa will get off-diagonal elements due to the RG running (here we present just the perturbative single leading logarithm),
\begin{eqnarray}
\delta Y_u \sim \frac{3}{32\pi^2} {\rm \ln}  \left(\frac{\Lambda}{\mu}\right) Y_d Y_d^\dagger Y_u(\Lambda) \ .
\end{eqnarray}
When we go to the mass eigenstate basis by diagonalizing $Y_u(\Lambda)+\delta Y_u$, this will affect the structure of the $\lambda$ matrix, which again leads to flavor violation in the up sector, such as $D-\bar{D}$ mixing. Similarly, the RG running can also generate non-zero $\lambda'$, which yields flavor violation in the down sector.
Fortunately, the relevant couplings are suppressed by both down type Yukawa couplings and the loop factor. Even for a very large $\Lambda$ of order the Planck scale we do not find any additional relevant  constraints with the above ansatz for the flavor structure at the scale $\Lambda$.

Because one of the diagonal elements of $\epsilon$ matrix can still be large, we explore here how LHC data can also be used to constrain $\epsilon_{ee}$ and $\epsilon_{\mu\mu}$. We find that it can already put a competitive constraint on these diagonal $\epsilon$'s to those from the LEP2 and NuTeV experiments.
For $\epsilon_{\tau\tau}$ the SuperK experiment provides the most important constraint.

The processes we study are charged-lepton pair production rates at high invariant mass at LHC, which can happen via a $t$-channel leptoquark exchange. The parton level process is $u\bar u \to e^+e^-$, or $\mu^+\mu^-$. For $\epsilon_{\alpha\alpha}$ larger than one percent and for Yukawa couplings no larger than 1, the leptoquark cannot be much heavier than a TeV. Therefore, the contact interaction analysis~\cite{Chatrchyan:2012hda} may not directly apply, and the bound in our case will be weaker.

We generate $e^+e^-$ (and $\mu^+\mu^-$) events in the leptoquark model using FeynRules~\cite{Christensen:2008py} and MadGraph~\cite{Alwall:2007st}, and compare them with
the 8\,TeV, 20.6\,${\rm fb}^{-1}$ LHC data given in~\cite{CMSmumu, CMSee}.
Non-observation of excess beyond SM background at LHC can be translated into upper bounds on $\epsilon_{ee}$ and $\epsilon_{\mu\mu}$, which are shown as the red curves in the first row of Fig.~\ref{LHC}. 
In the same plot, the blue region is the LHC exclusion on the mass of the leptoquark from its pair production and decays.
The yellow region is excluded by NuTeVneutrino scattering experiment.
The orange region is excluded by the Super-K experiment as mentioned in the introduction.
The LEP2 exclusion on $|\epsilon_{ee}|$ from $e^+e^-\to u \bar u$ channel is the green region.
The magenta region is non-perturbative regime when the leptoquark coupling $|\lambda_{\alpha i}|>\sqrt{4\pi}, \ (\alpha=e, \mu, \ i=u,d)$.
As we can read from the plot,  LHC data already sets a strong limit implying that $|\epsilon_{ee}|,\ |\epsilon_{\mu\mu}|$ should be less than a few percent. 
For the $|\epsilon_{ee}|$ case, it is a stronger constraint than LEP2 and NuTeV.
The LHC also already does better HERA~\cite{Abramowicz:2012tg} in the search for leptoquarks.
Future running of LHC will further improve the bound.

Note also that the production of single leptoquark and a lepton at LHC may provide relevant constraints~\cite{Belyaev:2005ew, FileviezPerez:2008dw}, especially when the leptoquark Yukawa coupling is large.

To summarize, given the present experimental constraints from the LHC and charged lepton flavor physics, we find the doublet leptoquark model  allows an interesting pattern for the flavor structure of the epsilon matrix. All the off diagonal elements of the $\epsilon$ matrix are constrained to be less than about $10^{-3}$, but $\epsilon_{ee}$ or $\epsilon_{\mu\mu}$ can be at the $10^{-2}$ level, while $\epsilon_{\tau\tau}$ can be at the 0.1 level.

\subsection{Another  $SU(2)_L$ Doublet Leptoquark Model}

Before closing this section, we consider briefly another model with an $SU(2)_L$ doublet leptoquark with hypercharge $-1/3$. The Yukawa couplings of $S$ to the quarks and leptons are given by,
\begin{equation}
\begin{split}
\mathcal{L} &= \lambda_{ij} \overline{L_i} P_R d_j S + {\rm h.c.} \\
&= \lambda_{ij} (\overline{\nu_i} P_R d_j X + \overline{e_i} P_R d_j Y) + {\rm h.c.} \ .
\end{split}
\end{equation}
The only difference from the model in Eq.~(\ref{lagrangeleptodoublet}) is that now neutrinos have new interactions with the down quark.
In the effective Hamilonian  language, integrating out $S$ yields $c_7\neq0$ and other coefficients vanishing in the Eq.~(\ref{L2Q2}).

In Fig.~\ref{LHC}, we have also shown the constraints  on  $|\epsilon_{ee}|$ and $|\epsilon_{\mu \mu}|$. The LHC limit in $d\bar d \to e^+e^-$, or $\mu^+\mu^-$ channels. 
Because of the relatively lower down quark PDF, the current LHC limit on $|\epsilon_{ee}|$ is of comparable order as the LEP2 and NuTeV limits, but we expect the future running of LHC will substantially improve the bound.
As in the previous model, all the off diagonal elements of the $\epsilon$ matrix are constrained to be less than about $10^{-3}$, but $|\epsilon_{ee}|$ or $|\epsilon_{\mu\mu}|$ can be at the $10^{-2}$ level, while $|\epsilon_{\tau\tau}|$ can be at the 0.1 level.

\subsection{Gauged $U(1)$ Models}

Here we briefly discuss  two models with an additional $U(1)$ gauged. One with the new gauge group $U(1)_{B-L}$ and the other with the new gauge group $U(1)_{L_e-L_{\mu}}$.
In both model, a right-handed neutrino has to be introduced for each family of SM fermions to cancel anomalies.

At leading order in perturbation theory, gauging $B-L$ gives the matching conditions
\begin{eqnarray}
\begin{split}
&c_1^{\alpha \beta} = {g^2 \over M_V^2} (\delta^{\alpha\beta} - \delta^\alpha_e \delta^\beta_e/2), \ \ \  c_3^{\alpha \beta}={g^2 \over M_V^2}\delta^{\alpha\beta}, \\
&c_4^{\alpha \beta} = c_6^{\alpha \beta} = c_7^{\alpha \beta}=-{g^2 \over 3M_V^2}\delta^{\alpha\beta} \ ,
\end{split}
\end{eqnarray}
where we choose $c_2^{\alpha\beta}=0$ to remove the redundancy, 
and the other $c_i$ are zero. Here $g$ is the gauge coupling and $M_V$ is the vector boson mass. From Eqs.~(\ref{epsL}) and (\ref{epsQ}), we get
\begin{equation}
\label{bl}
\epsilon_{ee}= \epsilon_{\mu \mu}=\epsilon_{\tau \tau}=-{g^2 \over \sqrt{2}{G_F} M_V^2} \ ,
\end{equation}
where the $\epsilon$'s not explicitly given are zero. 

On the other hand, gauging $L_e-L_{\mu}$ lepton number gives,
\begin{eqnarray}
\begin{split}
c_1^{\alpha \beta}&= {g^2 \over M_V^2} \left(\delta^{\alpha}_e \delta^\beta_e/2 - \delta^{\alpha}_\mu \delta^\beta_\mu \right) \ , \\
c_3^{\alpha \beta}&={g^2 \over M_V^2} \left(\delta^{\alpha}_e \delta^\beta_e - \delta^{\alpha}_\mu \delta^\beta_\mu \right) \ ,
\end{split}
\end{eqnarray}
and again we set $c_2^{\alpha\beta}=0$ to remove the redundancy.
This yields
\begin{equation}
\label{mutau}
\epsilon_{ee}= - \epsilon_{\mu \mu}= {g^2 \over \sqrt{2}{G_F} M_V^2} \ .
\end{equation}
In both models, we find the LEP2 bound on contact interactions~\cite{LEP:2003aa, Carena:2004xs} implies that $|\epsilon_{\mu\mu}|<0.8\times10^{-3}$.

\begin{figure}[t]
\vspace{-3cm}
\includegraphics[width=1.0\columnwidth]{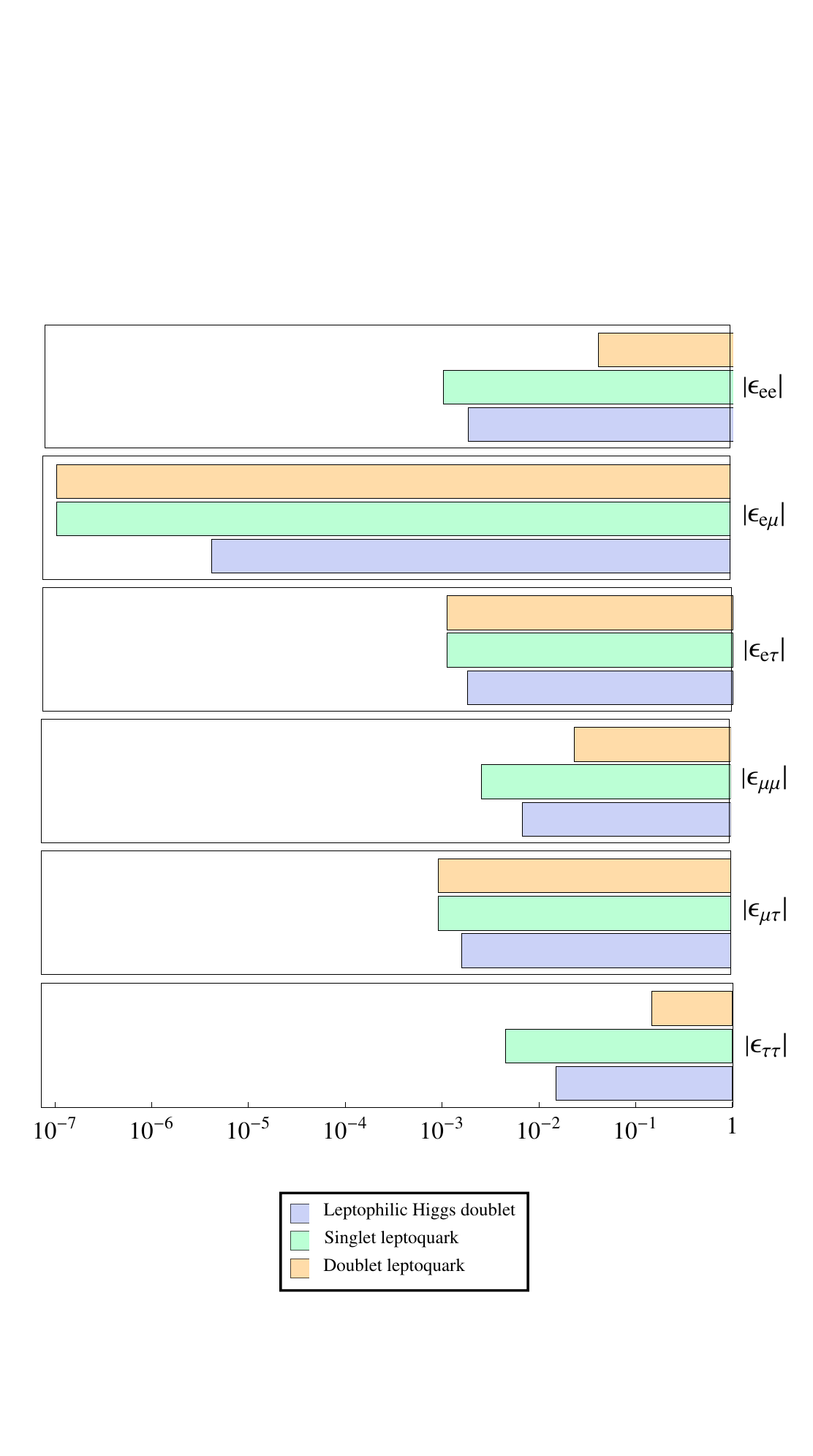}
\vspace{-2cm}
\caption{Summary of constraints on $|\epsilon_{\alpha\beta}|$ in various simple renormalizable BSM models for non-standard neutrino interactions. 
Shaded bands represent excluded values of epsilons for different models.
We would like to highlight that for the doublet leptoquark model one of the diagonal elements $\epsilon_{\alpha\alpha}$ can still be ``large'', {\it i.e.}, $\sim10^{-2}$ for $\epsilon_{ee}$, $\epsilon_{\mu\mu}$, and $\sim10^{-1}$ for $\epsilon_{\tau\tau}$.
}\label{summary}
\end{figure}

\section{Concluding Remarks}

In this paper we have considered several models and an operator analysis of new physics that contributes to neutrino interactions with matter. These interactions are characterized by the parameters $\epsilon_{\alpha \beta}$, where  $\alpha, \beta =e,\mu, \tau$. 

We  listed the set of relevant operators at dimension six and gave constraints on the Wilson coefficients from present data. Since the left handed neutrinos are in a doublet with the charged leptons, there are constraints from precision flavor physics and collider physics on the Wilson coefficients. 
Making the simplifying assumption that the matter was neutral ($n_e=n_p$) and had equal numbers of protons and neutrons ($n_p=n_n$) we expressed the elements of the matrix $\epsilon$ (that can be measured in future neutrino oscillation experiments {\it i.e.}, the off diagonal elements and the difference between diagonal ones) in terms of the Wilson coefficients of the operators.

We also analyzed what values of the elments of the matrix $\epsilon$ are allowed, given current experimental constraints, in various simple extensions of the SM that contain a new U(1) gauge boson or a new  scalar representation of the SM gauge group. Our analysis of  these extensions are meant to give the reader a sense what is the plausible range for the  $\epsilon_{\alpha \beta}$'s given present experimental constraints while the operator analysis informs on what is possible.  The model dependent results of our analysis are summarized in the Fig.~\ref{summary}, and are in a format that can easily be compared with the sensitivity of the proposed LBNE experiment (see Fig.~4-33 in Ref.~\cite{Adams:2013qkq}).
Models we discussed where the allowed values of the  $\epsilon_{\alpha \beta}$'s were all restricted to be no larger than about $10^{-4}$  are not presented in Fig.~\ref{summary} .  If non SM neutrino interactions were discovered in the future, the hierarchy amongst the $\epsilon$'s would help to differentiate amongst new physics models. 
In particular, in some of the leptoquark models we find that  one of the diagonal elements $\epsilon_{\tau\tau}$ can  be as large as $\sim0.1$, or one of $\epsilon_{ee}$, $\epsilon_{\mu\mu}\sim0.01$.

In all of the scalar models Eq.~(\ref{relationepsilons}) was satisfied. It relates the magnitude of the diagonal $\epsilon$'s to the off diagonal ones. However, this relationship was not satisfied by the models with an additional $U(1)$ gauge group.

Some the work done here has been discussed previously in the literature. However, there are a few novel aspects in our presentation of  the operator analysis. Furthermore, our discussion of the phenomenological aspects of the leptoquark models for neutrino oscillation physics is more complete than the previous literature. In particular, we studied how $t$-channel leptoquark exchange gives rise to $pp \to \ell{ \bar \ell}+X$ at the LHC and puts strong restrictions on neutrino interactions.

\bigskip
\section*{Acknowledgements}
We thank Michael Ramsey-Musolf, Ryan Patterson, Maurizio Pierini and Lisa Randall for useful discussions and correspondence.
This work is supported by the Gordon and Betty Moore Foundation through Grant \#776 to the Caltech Moore Center for Theoretical Cosmology and Physics, and by the DOE Grant DE-FG02-92ER40701.

\end{document}